\documentstyle[epsfig]{mn}

\def\ltsima{$\; \buildrel < \over \sim \;$}
\def\lsim{\lower.5ex\hbox{\ltsima}}
\def\gtsima{$\; \buildrel > \over \sim \;$}
\def\gsim{\lower.5ex\hbox{\gtsima}}
\def\wth{\omega(\theta)}

\def\om{\Omega_m}
\def\hth{\hat{\phi}}

\title[Halo Model Cross-Correlation Predictions]
{Quasar-Galaxy and Galaxy-Galaxy Cross-Correlations: Model Predictions 
with Realistic Galaxies}

\author[B. Jain, R. Scranton, R. K. Sheth]
{Bhuvnesh Jain$^1$,  
Ryan Scranton$^2$, Ravi K. Sheth$^2$ \\
$^1$ Department of Physics and Astronomy, University of Pennsylvania, 
Philadelphia, PA 19104 \\
$^2$ Department of Physics and Astronomy, University of Pittsburgh, 
     3941 O'Hara Street, Pittsburgh, PA 15260\\ 
}

\begin{document}
\maketitle

\begin{abstract}
Several measurements of QSO-galaxy correlations have reported signals
much larger than predictions of magnification by large-scale
structure. We find that the expected signal depends stronly on the
properties of the foreground galaxy population. On arcminute scales it
can be either larger or smaller by a factor of two for different galaxy
types in comparison with a linearly
biased version of the mass distribution. Thus the resolution of some of
the excess measurements may lie in examining the halo occupation 
properties of the galaxy population sampled by a given survey; this is
also the primary information such measurements will provide.  

We use the halo model of clustering and simulations to predict
the magnification induced cross-correlations 
and errors for forthcoming surveys. With the full Sloan Digital Sky 
Survey the statistical errors will be below 1 percent for the galaxy-galaxy
correlations and significantly larger for QSO-galaxy correlations. 
Thus accurate
constraints on parameters of the galaxy halo occupation distribution can be
obtained from small scale measurements 
and on the bias parameter from large scales. Since the lensing
induced cross-correlation measures the first moment of the halo occupation
number of galaxies, these measurements can provide the basis for interpreting
galaxy clustering measurements which measure the second and
higher order moments. 
\end{abstract}

\begin{keywords}
cosmology: dark matter --- cosmology: gravitational lensing
--- galaxies: clustering
\end{keywords}

\section{Introduction}

Gravitational lensing by large-scale structure along the line-of-sight
can alter the observed number density of galaxies on the sky (Gunn
1967).  Lensing increases the area of a given patch on the sky, thus
diluting the number density. On the other hand, galaxies fainter than
the limiting magnitude are brightened due to lensing and may be
included in the sample, thus increasing the number density. The net
effect, known as magnification bias, can go either way: it can lead to
an enhancement or suppression of the observed number density of
galaxies, depending on the slope $s$ of the number counts of galaxies
$N_0(m)$ in a sample with limiting magnitude $m$, $s= d\log N_0/dm$.
Magnification by amount $\mu$ changes the number counts to
(e.g. Broadhurst, Taylor \& Peacock 1995),
\begin{equation} 
N^\prime(m)=N_0(m) \mu^{2.5s-1} \, .
\label{mag1}
\end{equation} 	
For the critical value $s=0.4$, magnification does not change the 
number density; it leads to an excess for $s>0.4$, and a deficit
for $s<0.4$. 

This magnification effect also leads to an excess correlation of QSOs 
and foreground galaxies associated with the lensing mass distribution 
(Canizares 1981; Schneider, Ehlers \& Falco 1992; 
see also Keel 1982; Peacock 1982). 
It is measured through the excess or deficit in the counts of galaxies 
around background QSOs. It is critical to ensure
that the galaxies are not physically associated with the QSOs; this 
can be done by choosing appropriate cuts in the redshifts of the two
populations. Several measurements of the QSO-galaxy
cross-correlation have been made: significant excesses of galaxies
around QSOs have been found, although the magnitude of the effect measured 
remains well in excess of the expectations of dark-matter theory 
with simple biasing, as discussed below 
(Bartelmann \& Schneider 2001 give an excellent 
review of the theory and observations).  This has remained an unresolved
problem, manifested in particular in measurements on angular scales
of order an arcminute (see the discussion in Ben\'\i tez et al 2001). 
Calculations of the expected amplitude of this effect 
have been done using both linear and non-linear dark-matter clustering 
(Bartelmann 1995).  

Several observational measurements of QSO-galaxy correlations have
been made, beginning with the mid-1980's. 
While the results vary considerably, 
most studies with radio loud QSOs have found correlations in excess of lensing
predictions (Tyson 1986; Fugmann 1988,1990; Hammer \& Le F\'{e}vre 1990; 
Hintzen et al. 1991; Drinkwater et al. 1992;
Thomas et al. 1995; Bartelmann \& Schneider 1993b, 1994; Bartsch, Schneider,
\& Bartelmann 1997; Ben\'\i tez et al. 1995; Ben\'\i tez
\& Martinez-Gonzalez 1995, 1997; 
Norman \& Williams 1999; Norman \& Impey 1999, 2000). 
Although these results are qualitatively in agreement with the magnification 
bias effect, in most cases the amplitude of the correlation is much higher 
than that expected from gravitational lensing models.
Studies that have used optically selected QSOs
have found both positive and null or negative correlations. 
Webster et al (1988), Rodrigues-Williams \& Hogan (1991), 
Williams \& Irwin (1998) and Gaztanaga (2003) 
found positive correlations that are much
stronger than the predictions from the magnification bias effect, while
Boyle et al. (1988); Romani \& Maoz (1992); 
Ferreras, Ben\'\i tez \& Martinez-Gonzalez (1997) and 
Croom \& Shanks (1999) found null or even negative correlations, which 
in some cases cannot be explained by the lensing hypothesis. 
Some of the measured cross-correlations are likely to be affected
by the incompleteness of the QSO samples, and most have small
samples of galaxies and QSOs with little information on the redshifts
of the galaxy population. 

Even with these caveats, it is fair to say that there are severe discrepancies
between measured QSO-galaxy correlations and theoretical
predictions. Two kinds of resolutions have been discussed in the
literature: observational effects, such as bias in the selection
of QSO samples, physical associations of QSOs and galaxies, and
dust obscuration; and improved theoretical modeling of the lensing
(see the discussions in Ben\'\i tez, Sanz \& Martinez-Gonzalez 2001 
and Bartelmann \& Schneider 2001). On the theory side, various nonlinear 
effects have been modeled. Bartelmann \& Schneider (1993) and 
Bartelmann (1995) provided the framework for theoretical 
modeling in terms of large-scale structure parameters. Dolag \& Bartelmann
(1997) and Sanz et al (1997) made magnification predictions including nonlinear
gravitational evolution, as done by Jain \& Seljak (1997) for the shear,  
while Williams (2001) and M\'enard et al (2002) studied corrections to 
the weak lensing approximations. However, severe discrepancies remain; in
a careful analysis, Ben\'\i tez, Sanz \& Martinez-Gonzalez (2001) find 
a signal that is a factor of a few larger than model predictions on
scales of $1-5'$. 

In this paper we model the galaxy distribution using the halo model
of clustering (e.g. Cooray \& Sheth 2002). We compute QSO-galaxy
and galaxy-galaxy cross-correlation using this model for the galaxies
and compare it to the linear bias approach used so far in the literature. 
We seek to explain some of the discrepancies between measurements
and lensing predictions with our galaxy model. 
Under the lensing hypothesis, the QSO-galaxy cross-correlation is 
due to the cross-correlation of galaxy number density with magnification. 
This in turn can be expressed as the projection of the galaxy-mass
cross-correlation. We only modify the next, final step, of expressing the 
galaxy-mass cross-correlation as a linear bias factor times the mass-mass 
correlation. Instead we use the halo model of clustering to populate
halos of given mass with galaxies, and then compute the cross-correlation
consistently. One might expect that on small enough scales, $\lsim 1$ Mpc, 
the results would differ significantly from a linear bias model 
because the occupation of halos with galaxies is complicated and depends
strongly on galaxy type. Our approach to galaxy-mass correlations 
is similar to that of Seljak (2000), but there are differences in the 
galaxy model used. 

We aim our predictions at forthcoming survey data, primarily from
the Sloan Digital Sky Survey (SDSS), and the CFHT Legacy survey. 
We make predictions for the QSO-galaxy and the galaxy-galaxy
cross-correlation. 
Section 1 contains the formalism for the cross-correlation calculation. 
Section 2 details the halo model and the prescription for assigning
galaxies to halos. The results are presented in Section 4 and we conclude
in Section 5. 

\section{Formalism}

Angular correlations can be expressed as projections of the
3-dimensional power spectrum using Limber's equation.  We follow the
convention and derivations of Moessner \& Jain (1998).  Let
$n_1(\hth)$ be the number density of foreground galaxies with mean
redshift $\langle z_1\rangle$, observed in the direction $\hth$ in the
sky, and $n_2(\hth)$ that of the sample with a higher mean redshift
$\langle z_2\rangle > \langle z_1 \rangle $. The angular
cross-correlation function is then defined as:
\begin{equation} 
\wth=\left< \delta n_1(\hth) \delta n_2(\hth^\prime)   \right>  \; ,
\label{wcross}
\end{equation} 
where $\delta n_i(\hth) \equiv (n_i(\hth)-{\bar{n}_i})/{\bar{n}_i}$,
with ${\bar{n_i}}$ the average number density of the $i$th sample.
The measured $\delta n_i$ is the sum of fluctuations due to the true
clustering of galaxies $\delta n_i^g$, and due to magnification bias
$\delta n_i^{\mu}$.

The fluctuations on the sky due to intrinsic
clustering are a projection of the galaxy fluctuations along the
line-of-sight, weighted with the radial
distribution $W(\chi)$ of the galaxies,
\begin{equation} 
\delta n_i^g(\hth)= \int_0^{\chi_H} d \chi W_i(\chi) \delta^g(r(\chi)
\hth, a) \; , 
\end{equation}
where $a$ denotes the expansion scale factor. 
The comoving radial coordinate $\chi$ and the comoving angular
diameter distance $r(\chi)$ are explicitly defined in e.g. Jain \& Seljak
(1997). $\chi_H$ is the distance to the horizon.

In the weak lensing limit the magnification is $\mu=1+2 \kappa$, where
the convergence $\kappa$ is a weighted projection of the density field
along the line-of-sight (see equation \ref{kappa1} below; see also 
M\'enard et al 2002 for tests of the weak lensing limit). Equation \ref{mag1} 
then gives the magnification induced fluctuations in number counts as
\begin{equation} 
\delta n_i^{\mu}(\hth) = 5(s-0.4)\kappa_i(\hth),
\label{Np}
\end{equation}
with the convergence $\kappa$ given by
\begin{equation} 
\kappa_i(\hth)=\frac{3}{2} \om \int_0^{\chi_H} d \chi\ g_i(\chi)
\frac{\delta(r \hth,\chi)}{a} \; ,
\label{kappa1}
\end{equation}
where the radial weight function $g(\chi)$ can be expressed in terms
of $r(\chi)$; for a delta-function distribution of source galaxies at
$\chi_S$ it is $g(\chi)=r(\chi)r(\chi_S-\chi)/r(\chi_S)$.

The cross-correlation $\omega(\theta)$ is then given by the sum of
four terms. In the case where the background and foreground samples have no
overlap, the cross-correlation is dominated by (Moessner \& Jain 1998),
\begin{eqnarray}
\omega (\theta)&=& 3 \om (2.5 s_2-1) 4 \pi^2 \int_0^{\chi_H} d \chi
W_1(\chi) \frac{g_2(\chi)}{a} \nonumber \\
&& \times \int_0^\infty dk\, k\, P_{\rm gm}(\chi, k)\,
J_0\left[k r(\chi)\theta\right] \; , 
\label{omegagl}
\end{eqnarray}
where $P_{\rm gm}(\chi, k)$ is the galaxy-mass cross-power spectrum, 
Note that the expression for the mean tangential shear around a
foreground galaxy is identical to the above equation except for the
replacement of $J_0$ by $J_2$ (this follows from the relation
$\gamma_t(\theta)=-1/2 \ d\, \bar\kappa(\theta)/d\, {\rm ln} \theta$). 
The difference in Bessel functions that appear for the shear and
magnification can be useful in combining information from the two
measurements. They weight length scales in different ways because
the magnification measures probe the local projected density while
the shear measures probe the integrated mass inside an aperture. Thus
with comparable signal to noise measurements from the two probes, one can 
gain not only an imrovement in overall sensitivity, but also probe a larger
range of length scales. 

\section{Models for galaxy-dark matter correlations}

We express galaxy-mass cross-correlations by developing a halo
model based description of the cross-power spectrum, $P_{\rm gm}(\chi, k)$, 
introduced above. $P_{\rm gm}$ can be modeled in at least three ways: 
(i) Take it to be a bias factor times the matter-matter power spectrum; 
(ii) Assume there is a one-to-one correspondence between galaxies and 
     halos, and that the galaxy sits at the centre of its parent halo;  
(iii) Use the halo model (Seljak 2000; Peacock \& Smith 2000; 
    Scoccimarro et al. 2001; see the recent review by Cooray \& Sheth
    2002), which allows for the possibility that the number of
    galaxies in a halo may vary from halo to halo, and allows for 
    the possibility that the distribution of galaxies around the 
    parent halo centre may depend on galaxy type.  

In the halo model, all dark matter particles and galaxies inhabit 
halos.  This means that 
\begin{eqnarray}
P_{\rm gm}(k) &=& P_{\rm gm}^{\rm 1h}(k) + P_{\rm gm}^{\rm 2h}(k)\nonumber\\
P_{\rm gm}^{\rm 1h}(k) &=& \int {\rm d}m\,n(m)\,{m\over\bar\rho}\,
                  {\langle N_{\rm gal}|m\rangle\over\bar n_{\rm gal}}\ 
                   |u(k|m)|\,|u_{\rm gal}(k|m)| \nonumber\\
P_{\rm gm}^{\rm 2h}(k) &\approx& P_{\rm lin}(k)\ 
\int {\rm d}m\,n(m)\,{m\over\bar\rho}\,b(m)\,u(k|m) \nonumber\\
&& \times \int {\rm d}m\,n(m)\,{\langle N_{\rm gal}|m\rangle\over\bar 
n_{\rm gal}}\,       b(m)\,u_{\rm gal}(k|m) \nonumber\\
\bar\rho &\equiv &  \int  {\rm d}m\,n(m)\,m \nonumber\\ 
\bar n_{\rm gal} & \equiv & 
\int\,\, {\rm d}m\,n(m)\,\langle N_{\rm gal}|m\rangle.
\label{noCD}
\end{eqnarray}
Here $n(m)$ is the comoving number density of halos which have mass 
$m$, $\langle N_{\rm gal}|m\rangle$ is the average number of galaxies 
in halos of mass $m$, $b(m)$ is the bias parameter, 
and $u(k|m)$ and $u_{\rm gal}(k|m)$ are the 
Fourier transforms of the density run of dark matter and galaxies around 
the centres of the halos they inhabit.  The models assume that the 
density run can be well parametrized by allowing for a depedence on 
halo mass:
\begin{equation}
u(k|m) = {1\over(2\pi)^3}
{\int {\rm d}r\,4\pi r^2\,\rho(r|m)\,\sin(kr)/kr \over
 \int {\rm d}r\,4\pi r^2\,\rho(r|m)},
\end{equation}
where the factor of $(2\pi)^{-3}$ defines our Fourier transform 
convention, and the integral in the denominator defines the mass $m$ 
of the halo.  The denominator normalizes $(2\pi)^3 u(k|m)$ so that it is unity 
at small $k$; for all profiles of interest, it decreases to zero as 
$k$ increases, although the decrease need not be monotonic.  

Option (ii)---one and only galaxy per halo, and the galaxy sits at the 
halo centre---corresponds to setting $N_{\rm gal}(m)=1$
and $(2\pi)^3 u_{\rm gal}(k|m)=1$ for all $m$ (we use $N_{\rm gal}(m)$ 
to denote $\langle N_{\rm gal}|m\rangle$ for simplicity). This is
a reasonable model for the Large Red Galaxy (LRG) sample provided
by the SDSS; results for this model are shown in Figure \ref{fig:halo}. 

Option (iii) requires additional assumptions to fully specify the 
galaxy distribution.  For example, one might expect the distribution 
of galaxies around the halo centre to depend on galaxy type.  
In what follows, we will assume for simplicity that the galaxies 
in a halo trace the dark matter:  therefore, we set 
$u_{\rm gal}(k|m)=u(k|m)$.  Sheth et al. (2001) show that this is 
a reasonable approximation.  (See Scranton 2001 for more discussion 
of how $u_{\rm gal}$ might depend on galaxy type).  

It is probably accurate to assume that one of the $N_{\rm gal}$ galaxies 
in a halo sits at the halo centre -- this is almost certainly a good 
assumption if there is only one galaxy in a halo.  
To account for this, one assumes that the other galaxies get the usual 
density profile denoted by $u$, whereas the central galaxy is simply unity.  
To see the effect of this on the two-halo term, we must average both 
pieces of the two halo term over $p(n|m)$, the probability of having
$n$ galaxies in a halo of mass $m$, with the requirement that $n>0$.
With this averaging, we will replace $N_{\rm gal}(m) u_{\rm gal}(k|m)$ 
by an effective number of galaxies, $N_{\rm eff}$, which
is a function of $k$ for given $m$. It is given by (Cooray \& Sheth 2002)
\begin{eqnarray}
 N_{\rm eff}(k|m) &\equiv & \sum_{n>0} \ \left[ 1 + (n-1)\ u(k|m)\right] \ 
p(n|m) \nonumber\\
&+& \sum_{n>0} \ \langle n-1|m\rangle\,u(k|m) + u(k|m)\,p(0|m) . 
\end{eqnarray}
The first term can be written as $[1 - p(0|m)]$, allowing us to 
express the above equation as
\begin{equation}
 N_{\rm eff}(k|m) = [1 - p(0|m)]\,[1 - u_{\rm gal}(k|m)] 
 + \langle n|m\rangle\, u_{\rm gal}(k|m) .
\end{equation}
Since both factors in the first term are positive, 
there is an enhancement in power  from always placing one 
galaxy at the halo centre.  Since $u(k|m)$ decreases as $k$ increases, 
the enhancement in power is largest on small scales (large $k$).  
In sufficiently massive halos one might expect to have many galaxies, 
and so $p(0|m)\ll 1$.  In this limit, the expression above becomes 
 $1-u(k|m) + \langle n|m\rangle\,u(k|m) = 1+\langle n-1|m\rangle\,u(k|m)$.  
On the other hand, if most halos have no galaxies, then $p(1|m)$ is 
probably much larger than all other $p(n|m)$ with $n\ge 2$.  Then the 
leading order term in the sum above is $p(1|m)$.  Since 
$\langle n|m\rangle \equiv \sum np(n|m) \approx p(1|m)$, we have that 
$N_{\rm eff}(k|m)\approx\langle n|m\rangle$.  In this limit, only a 
fraction $\langle n|m\rangle\ll 1$ of the halos contain a galaxy, and 
the galaxy sits at the halo centre, so there is no factor of $u$.  

The contribution of the galaxy counts to the one halo term of the 
galaxy--mass correlation function is similar.  Using the expressions 
above yields 
\begin{eqnarray}
 P_{\rm gm}^{\rm 1h}(k) &=& \int {\rm d}m\,n(m)\,{m\over\bar\rho}\,|u(k|m)|
  \, {N_{\rm eff}(k|m)\over \bar n_{\rm gal}} \nonumber\\
 P_{\rm gm}^{\rm 2h}(k) &\approx& P_{\rm lin}(k)\ 
 \int {\rm d}m\,n(m)\,{m\over\bar\rho}\,b(m)\,u(k|m)\ \nonumber\\
 && \qquad \times \int {\rm d}m\,n(m)\,b(m)\, {N_{\rm eff}(k|m)\over \bar n_{\rm gal}} .
\label{pkmg}
\end{eqnarray}

If the run of galaxies around the halo centre is not the same as of the 
dark matter, then one simply uses $u_{\rm gal}$ instead of $u$ in 
$N_{\rm eff}$.  If the two-halo term usually does not dominate the 
power on small scales (this is almost always a good approximation), it 
is reasonable to ignore the enhancement in power associated with the central 
galaxy, and to simply set 
 $N_{\rm eff}(k|m)\approx \langle n|m\rangle\, u_{\rm gal}(k|m)\approx 
 \langle n|m\rangle$.  The one-halo term requires knowledge of $p(0|m)$.  
Since $p(0|m)$ is usually unknown, some authors (e.g., Seljak 2000) 
interpolate between the two limits discussed earlier by setting 
$N_{\rm eff}=\langle n|m\rangle\, u(k|m)$ if $\langle n|m\rangle \ge 1$, 
and $N_{\rm eff}=\langle n|m\rangle$ if $\langle n|m\rangle < 1$.  
It is worth noting that the one halo term 
of the galaxy--galaxy correlation function is only slightly more complicated, 
as shown in Cooray \& Sheth (2002).  

\subsection{Halo model details}

The halo model is specified by the mass function $n(m)$, bias $b(m)$
and the halo profile $u(k|m)$. For the first two functions we use
(e.g. Cooray \& Sheth 2002)
\begin{eqnarray}
 n(m,z)\,{\rm d}m &=&
    {\bar\rho\over m}\,\nu f(\nu)\,{{\rm d}\nu\over\nu}, \nonumber \\
\nu f(\nu) &=& A(p)\left(1 + (q\nu)^{-p}\right) 
\,\left({q\nu\over 2\pi}\right)^{1/2}\,\exp\left(-{q\nu\over 2}\right); 
\nonumber \\
b(m) &=& 
1 + {q\nu-1\over\delta_{\rm sc}(z)} + {2p/\delta_{\rm sc}(z)\over
1+(q\nu)^p}, \nonumber \\
{\rm where}\quad \nu & \equiv & {\delta^2_{\rm sc}(z)\over \sigma^2(m)},
\label{nmfv}
\end{eqnarray}
and $p\approx 0.3$,
$A(p) = [1 + 2^{-p}\Gamma(1/2-p)/\sqrt{\pi}]^{-1} \approx 0.3222$, 
and $q\approx 0.75$ (Sheth \& Tormen 1999).  
Here $\delta_{\rm sc}(z)$ is the critical density required for 
spherical collapse at $z$, extrapolated to the present time using 
linear theory, and 
\begin{equation}
\sigma^2(m) = {4\pi\over (2\pi)^3}\int_0^\infty 
{{\rm d}k\over k}\,k^3P_{\rm Lin}(k)\ W^2(kR_0),
\end{equation}
where $W(x)=(3/x^3)[\sin(x) - x\cos(x)]$ and $R_0=(3m/4\pi\bar\rho)^{1/3}$.  
That is to say, $\sigma(m)$ is the rms value of the initial fluctuation 
field when it is smoothed with a tophat filter of comoving size $R_0$, 
extrapolated using linear theory to the present time.  
If $p=1/2$ and $q=1$, then $n(m)$ is the same as that 
first written down by Press \& Schechter (1974), and $b(m)$ is the 
same as that given by Cole \& Kaiser (1989) and Mo \& White (1996).

In addition, we will assume that the halo profiles have the form 
given by Navarro, Frenk \& White (1996), truncated at the virial 
radius $r_{\rm vir}$ which is defined by requiring that 
$m = 4\pi/3 r_{\rm vir}^3\bar\rho\,\Delta_{\rm vir}$.  
For spatially flat universes with $\Omega_0=(1,0.3)$ and 
$\Lambda=1-\Omega$, $\Delta_{\rm vir}= (178,340)$.
The Fourier transform of the density run around a halo of mass 
$m$ is 

\begin{eqnarray}
u(k|m) &=& f(c) \ 
\Big[ \sin \kappa \Big( {\rm Si}[\kappa(1+c)]- {\rm Si}(\kappa) \Big)
\nonumber \\
&+& \cos \kappa \Big( {\rm Ci}[\kappa(1+c)]- {\rm Ci}(\kappa) \Big) -
\frac{\sin (\kappa c)}{\kappa (1+c)} \Big], 
\label{unfw}
\end{eqnarray}

where $f(c) = 1/[\ln(1+c)-c/(1+c)]$, 
$\kappa \equiv kr_{\rm vir}/c$, 
${\rm Si}(x) = \int_0^x {\rm d}t\,\sin (t)/t$ is the sine integral and 
${\rm Ci}(x) = - \int_x^\infty {\rm d}t\,\cos (t)/t$ is the cosine integral 
function. 
The concentration parameter of the halos depends on halo 
mass;  we use the parametrization of this dependence 
given by Bullock et al. (2001):  

\begin{equation}
 c(m) \approx 9 \left( {m\over m_*} \right)^{-0.13}.
\end{equation}

\subsection{Galaxy Model Details}

Our model for the galaxy distribution is taken from the GIF N-body 
simulations of the $\Lambda$CDM model, coupled to a semi-analytic galaxy 
formation model (Kauffmann et al. 1999).  Catalogs of galaxy 
positions, apparent and absolute magnitudes, colors and star-formation 
rates, at a range of redshifts, are available at 
{\tt http://www.mpa-garching.mpg.de/GIF/}.  
From the catalogs at redshifts $z=0.06$, 0.13, 0.27, 0.35, 
0.42, and 0.52 we selected apparent magnitude limited samples 
which satisfied $17<r^*<21$.  At each redshift we then computed 
the first and second factorial moments of the number of galaxies 
as a function of halo mass.  The parameters for the first moment 
are given in Table~\ref{tab:ngal}.  These galaxies also form a surface 
defined by the first and second factorial moments as a function of 
halo mass and galaxy absolute magnitude which we can also integrate
over given our apparent magnitude limits and a redshift (Scranton 2002). 

\begin{figure}
\vspace{13cm}
\caption{The mean number of galaxies as a function of halo mass, 
$N_{\rm gal}(m)$, is shown for a range of redshifts. The symbols 
show the measurements from GIF simulations, and the solid curves
shows our fits in Table 1. 
}
\includegraphics{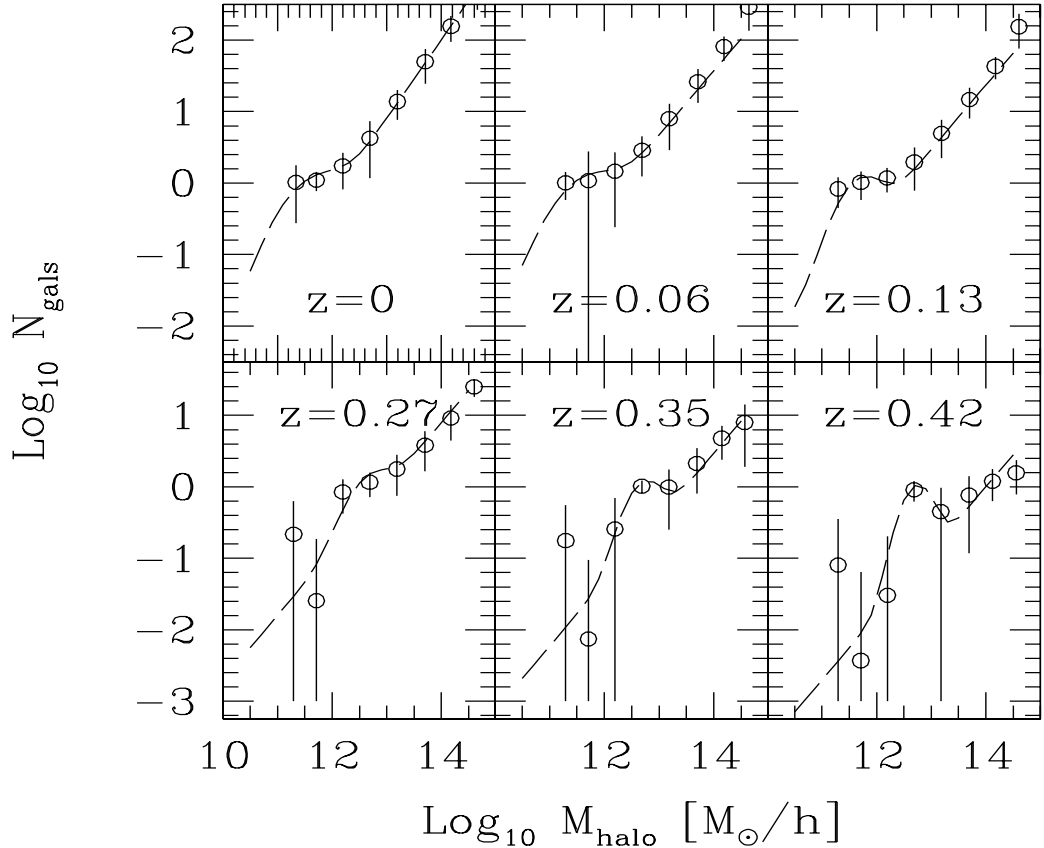}
\label{fig:ngm}
\end{figure}

In the GIF models, the mean number of galaxies in a halo 
$\langle n|m\rangle$ evolves little out to $z\sim 0.5$ if the 
galaxies have the same fixed rest-frame luminosity.  
Thus at fixed luminosity the evolution of galaxy 
clustering is driven by the evolution of the halo population.  
However, this means that at fixed apparent magnitude, the case 
in which we are most interested, $\langle n|m\rangle$ evolves more 
strongly.  This is because only the most luminous galaxies from the 
higher redshift samples satisfy the apparent magnitude limit, and, in 
halos which host only one galaxy, there is a tight correlation between 
halo mass and galaxy luminosity.  In principle, fits to 
$\langle n|m\rangle$ at fixed luminosity, along with the assumption 
that these relations do not evolve, allow one to estimate how 
$\langle n|m\rangle$ depends on redshift for the apparent magnitude 
limits we apply.  In practice, we chose the more tedious but accurate 
method of making redshift dependent fits to the apparent magnitude 
$\langle n|m\rangle$ relation in the GIF model.  
The results are shown in Figure~\ref{fig:ngm}. 

\begin{table}
\caption{Parameters which specify the mean number of galaxies 
in halos as a function of halo mass:  
  $N_{\rm gal}(m) \equiv (m/M_0)^\alpha + 
  A \exp\left [ -A_0 \left ( log_{10}(m) - M_B \right)^2 \right]$. 
This Table gives the parameter values for the qso-galaxy
case, with the foreground galaxies selected in 
the magnitude range $17<r'<22$. 
}
\centering
\begin{tabular}{cccccc}
 \hline 
 $z$ & log $M_0/h^{-1}M_\odot$ & $\alpha$ & $A$ & $A_0$ & 
$M_B/ \log \left ( h^{-1}M_\odot \right )$ \\
 \hline 
 0.06 & 12.30 & 1.05 & 0.98 & 11.56 & 11.55 \\
 0.13 & 12.47 & 1.03 & 0.96 & 0.62 & 11.42 \\
 0.27 & 12.74 & 0.94 & 0.87 & 1.65 & 12.10 \\
 0.35 & 12.82 & 0.79 & 0.84 & 7.41 & 12.25 \\
 0.42 & 13.11 & 0.76 & 0.91 & 7.3 & 12.45 \\
 0.52 & 13.67 & 0.84 & 1.03 & 10.07 & 12.68 \\
 \hline 
\end{tabular}
\label{tab:ngal}
\end{table}

\begin{table}
\caption{ $N_{\rm gal}(m)$ as in Table \ref{tab:ngal}, 
but for the galaxy-galaxy case, with the foreground galaxies
in the apparent magnitude range $17<r'<21$.   
}
\centering
\begin{tabular}{cccccc}
 \hline 
 $z$ & log $M_0/h^{-1}M_\odot$ & $\alpha$ & $A$ & $A_0$ & 
$M_B/ \log \left ( h^{-1}M_\odot \right )$ \\
 \hline 
 0.06 & 12.26 & 0.9 & 1.0 & 2.0 & 11.75 \\
 0.13 & 12.48 & 0.9 & 1.0 & 4.0 & 11.75 \\
 0.27 & 13.0 & 0.9 &  1.0 & 4.0 & 12.75 \\
 0.35 & 13.48 & 0.9 & 1.0 & 6.0 & 12.75 \\
 0.42 & 14.0 & 0.9 &  1.0 & 8.0 & 12.75 \\
 0.52 & 14.3 & 0.9 &  0.2 & 8.0 & 12.75 \\
 \hline 
\end{tabular}
\label{tab:ngal2}
\end{table}

\begin{table}
\caption{$N_{\rm gal}(m)$ as in Table \ref{tab:ngal2}, 
but for ``red'' foreground
galaxies, defined by the criterion $g'-r' > 0.65$. The functional 
form is $N_{\rm gal}(m) \equiv (m/M_0)^\alpha \exp(-M_1/m)$. 
}
\centering
\begin{tabular}{cccccc}
 \hline 
 $z$ & log $M_0/h^{-1}M_\odot$ & $\alpha$ & log $M_1/h^{-1}M_\odot$ \\
 \hline 
 0.06 & 11.70 & 1.2 & 10.0 \\
 0.15 & 11.70 & 1.0 & 11.7 \\
 0.26 & 12.18 & 0.8 & 12.0 \\
 0.35 & 12.18 & 0.5 & 12.0 \\
 0.42 & 12.85 & 0.5 &  12.0 \\
 0.52 & 12.85 & 0.5 &  12.0 \\
 \hline 
\end{tabular}
\label{tab:ngal3}
\end{table}

\subsection{Lens and source redshift distributions}

Magnification bias with galaxy-galaxy lensing should be detected with
SDSS data and forthcoming datasets such as the CFHT Legacy survey. 
The SDSS photometric survey will provide of order 1 million foreground
galaxies and 100 million background galaxies with photometric redshifts. 
The expected signal is sensitive to the mean redshifts of the foreground
and source populations, but only at 
the 10\% level or less to the shape of the distribution. Given that there is
some freedom in the selection of the two populations with photometric
redshifts, we have chosen simple distributions for our
model predictions. For a CFHT Legacy survey kind of dataset, we choose
the foreground population to be uniformly distributed over $0.2<z<0.5$
and the sources at $z=1$. For predictions for the SDSS survey, we choose 
a much closer foreground population with $0.1<z<0.2$, and the sources
at $z=0.4$. We have checked that for a realistic, broad source redshift 
distribution with the same mean redshift, the signal would be lower by 
about $5$\%. 

Our expected distributions of galaxies and QSOs from an SDSS-like dataset 
for measurement of the QSO magnification bias are much more constrained than 
for the galaxy-galaxy lensing case.  First, we expect that the efficiency of
photometric selection of QSOs, while high compared to previous work, will
force the use of spectroscopically-confirmed QSOs.  This will result in 
approximately $10^5$ QSOs, down three orders of magnitude from the number of 
galaxies in the photometric sample.  
Hence we will need to maximize the number of objects in our 
distributions rather than restricting them to narrower, easier-to-model bands 
in redshift.  
For galaxies, this means going to a broad cut in apparent magnitude: 
$17 < r^\prime < 22$, for example.  Using the luminosity function from the 
CNOC2 survey (Lin et al 1999), we can translate such an apparent magnitude cut 
in the SDSS filters into a redshift distribution (Dodelson et al 2002) well 
fit by a simple functional form:
\begin{equation}
\frac{dN}{dz} \propto z^a \exp \left [- \left ( z/z_0 \right)^b \right ].
\label{eq:galaxy-dndz}
\end{equation}

The QSO sample distribution is more complicated.  Due to a combination of 
the QSO luminosity function and the evolution of QSO colors in the SDSS 
filters with redshift, the redshift distriubtion of QSOs from the main SDSS
sample requires two functions of the form given in 
Equation~\ref{eq:galaxy-dndz}, with an appropriate amplitude ($A$) for each to 
properly characterize the relative abundance of QSOs below $z = 2$ and those 
above.  Table~\ref{tab:dndz} gives the parameters for the galaxy and QSO
redshift distributions, with the normalization set to $A=1$ for the galaxy
distribution.  In addition to these smooth distributions, we impose
a strict redshift cut of $z < 0.8$ for the galaxy distribution and $z > 1$ for
the QSO distribution to avoid intrinsic clustering between the two samples.

\begin{table}
\caption{Parameters for redshift distribution functions of galaxies and QSOs
  of the form $dN/dz = A z^a \exp[-(z/z_0)^b]$.}
\centering
\begin{tabular}{ccccc}
 \hline 
 Object & $A$ & $a$ & $z_0$ & $b$ \\
 \hline 
   Galaxies   & $1$  & $1.56$ & $0.296$ & $1.76$ \\
   Low-$z$ QSOs & $253.28$ & $2.58$ & $1.67$ & $13.44$ \\
   High-$z$ QSOs & $6.5 \times 10^{-4}$ & 9.13 & 3.35 & 18.37 \\ 
 \hline 
\end{tabular}
\label{tab:dndz}
\end{table}

\section{Results}

\begin{figure}
\vspace{16cm}
\caption{(i) The left panels show the angular cross correlation
computed using the linear mass power spectrum (lower dotted curves), 
the nonlinear mass power spectrum (upper dotted), 
and three galaxy models
(blue, total, red from bottom to top). The foreground galaxies are taken
to lie between $0.2<z<0.5$ ($0.1<z<0.2$) and the sources at $z=1$ ($0.4$) 
for the upper (lower) panels.  The error bars have been multiplied by 10 
(see discussion in the text). (ii)
The right panels show the ratio of the galaxy cross-correlations
to the PD prediction. On large scales it is reasonable to interpret
this ratio as a bias parameter, but on small scales it is important to
consider the details of the halo occupation distribution of galaxies. 
}
\includegraphics{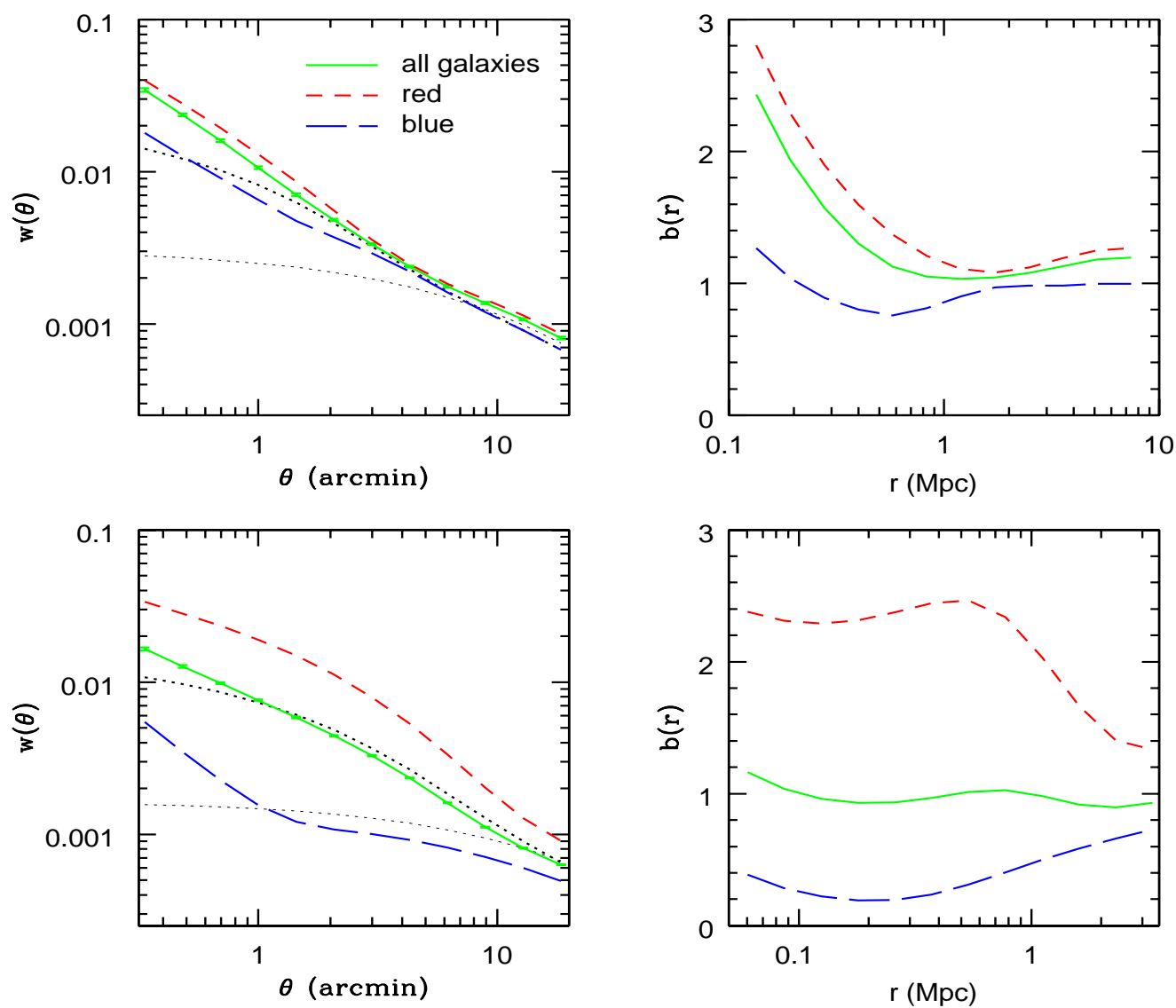}
\label{fig:gg}
\end{figure}

\begin{figure}
\vspace{8.5cm}
\caption{The angular cross correlation as in Figure \ref{fig:gg}, 
but for the qso-galaxy parameters appropriate for the SDSS survey. 
}
\includegraphics{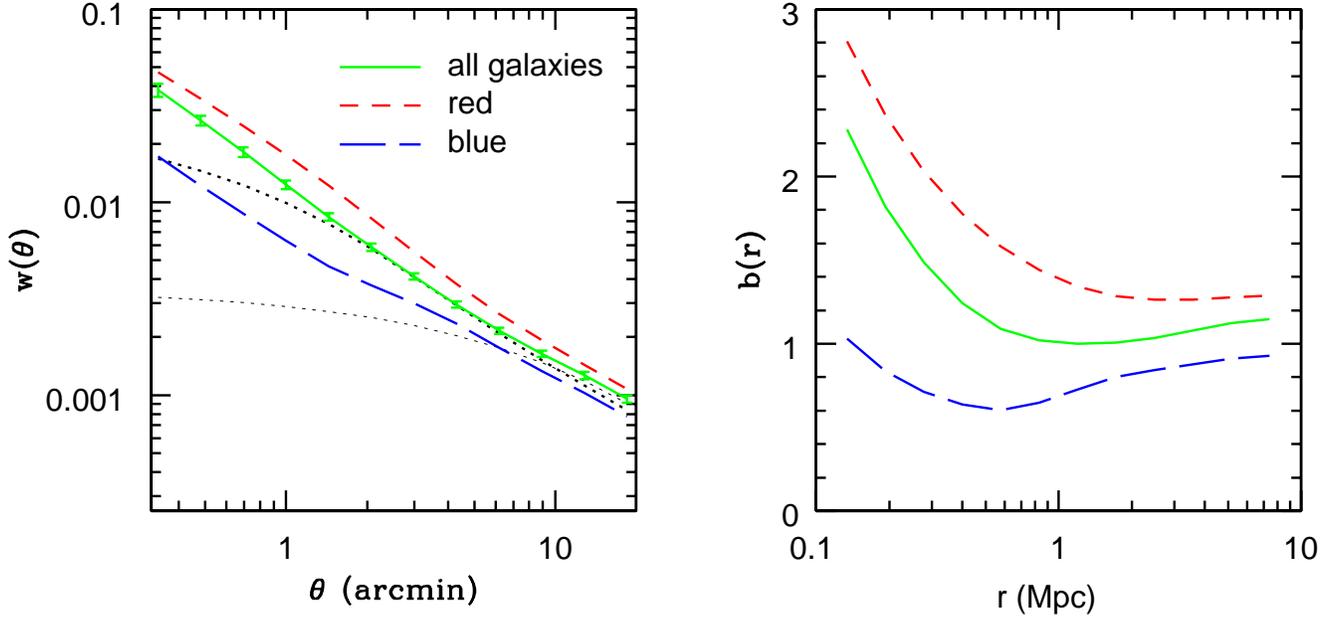}
\label{fig:qg}
\end{figure}

\begin{figure}
\vspace{8.5cm}
\caption{The angular cross correlation predictions for halos of masses 
$10^{10}, 10^{11}, 10^{12}, 10^{13} M_\odot $ (bottom to top), compared 
to the linear and PD predictions (solid curves). 
}
\includegraphics{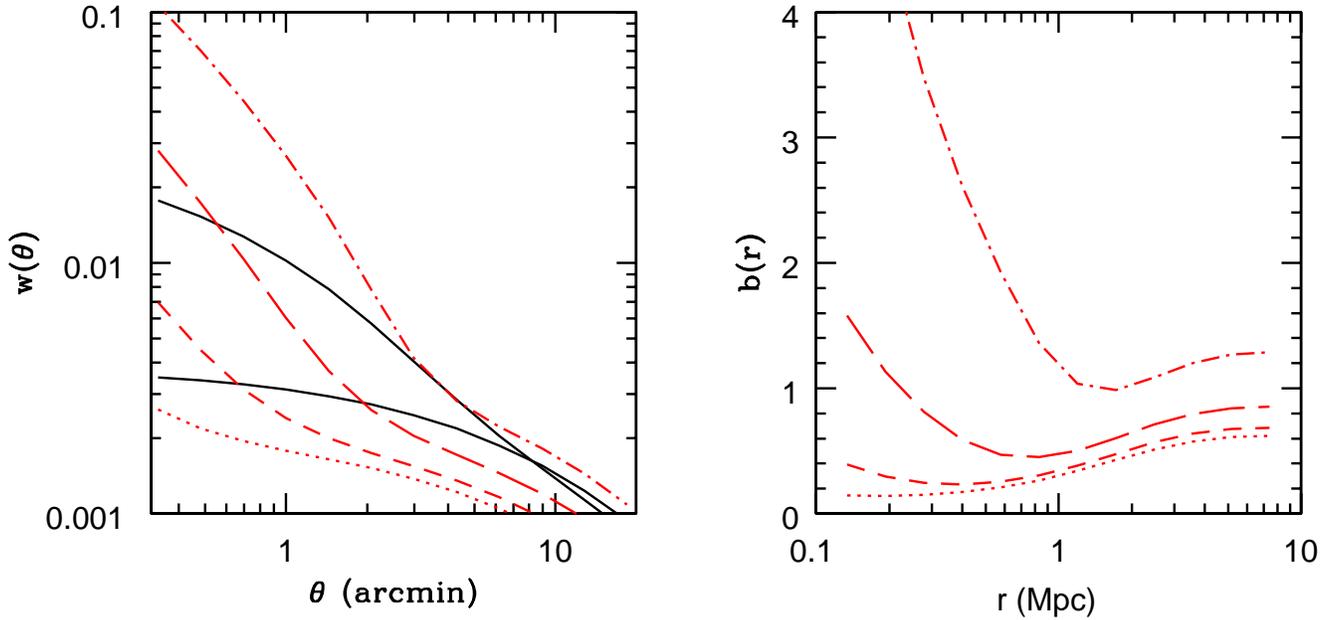}
\label{fig:halo}
\end{figure}

Figures \ref{fig:gg}-\ref{fig:halo} show the predicted cross-correlation 
for different foreground-background populations. We use equation 
\ref{omegagl}, with an $\Omega_{\rm m}=0.3, \Omega_{\lambda}=0.7$, 
$\Gamma = 0.21$ cosmological model with $\sigma_8=0.8$ and the galaxy-mass 
power spectrum determined using the halo model ingredients described in 
Section 3.  While $\sigma_8=0.8$ is close to the most recent determinations 
from the CMB and other datasets, there is uncertainty at about the 10\% level
in this parameter (e.g. Wang et al 2002; Spergel et al 2003; 
Verde et al 2003). Our results, 
especially for the higher redshift lens galaxies, are sensitive to $\sigma_8$:
for higher $\sigma_8$, the lens galaxies are less biased and vice versa.  
Note that we have left out the $(2.5 s - 1)$ factor which depends on 
galaxy type. It will modulate both the amplitude and possibly the sign 
of our plotted cross-correlation. 

We consider galaxy-galaxy
correlations in Figure \ref{fig:gg} for redshift distributions appropriate
to the CFHT Legacy survey in the upper panels and the SDSS survey
in the lower panel. The right panels show the ratio of the cross-correlation
to the Peacock-Dodds prediction. On large scales this reduces to the 
usual bias parameter, but on small scales it must be interpreted more
carefully in terms of properties of the halo occupation distribution. 
The results show that the cross-correlations are strongly dependent on
galaxy-type and angular scale. This arises from the modeling of the ``red''
and ``blue'' galaxy sub-samples that make up the full foreground sample. 
The galaxy samples are defined by the $N_{\rm gal}(m)$ models given in
Tables 2 and 3 for the full sample and for red galaxies (the blue galaxies
are given by the difference of the two). Similar trends are seen in 
Figure \ref{fig:qg}, which uses the galaxy distribution in Table 4 and
a mean redhisft $z=1.75$ for the quasars. 

It is interesting that for red galaxies, lowering the redshifts of the 
source and lens population does not change the signal significantly 
(comparing the upper and lower panels of Figure \ref{fig:gg}). 
This is a consequence of how little $N_{\rm gal}(m)$ evolves 
between $z=0.5$ and $z=0$ for these galaxies. The net change in lensing
signal as one lowers the redshifts is a trade off between two opposing 
effects: a reduction due to smaller lensing path length, and an increase 
because a given angle corresponds to a smaller length scale at the lens 
redshift. With a simplified power law model for the galaxy-mass power 
spectrum one can see that the net signal can increase with decreasing 
redshift. Let $P_{\rm gm}(k,a) \propto 
k^{-\alpha} a^\beta$, and let the lenses and sources be at a single
redshift such that the distance to sources is twice the distance to 
the lenses. Then for an Einstein-de Sitter model equation \ref{omegagl}
gives the analytic scaling
\begin{equation}
w(\theta)\propto a_{\rm lens}^{-{3\over 2}+{\alpha\over 2}+\beta} \ 
\theta^{2-\alpha}. 
\label{scaling}
\end{equation}
Thus for $\alpha=1$ (corresponding to a two-point correlation function
with logarithmic slope of $-2$), one can see that $\beta > 1$ would
lead to an increasing signal with decreasing lens redshift. 
The choice $\alpha=1$ and  $\beta=1$ is in 
good agreement with the behavior of the red galaxy sample, but not of
the blue galaxies for which the cross-correlation is lower for lower 
redshifts of the lenses and sources. The differences
between red and blue galaxies arise primarily from the difference in the
$N_{\rm gal}(m)$ relation for them (Sheth et al 2001; Scranton 2002). 
On larger scales, corresponding to 
$r > 1$Mpc, the difference between predictions for red and blue 
galaxies decreases. In the linear regime they
are both expected to follow the growth rate of mass fluctuations, so the
large-scale differences are primarily due to differences in $\alpha$. 

We estimate the contribution of Poisson errors and sample variance 
on the measured $\omega(\theta)$. Scranton (2002) discusses 
the relative contributions of these errors and of the 
Gaussian and non-Gaussian terms in the covariance. Over the scales
we have considered, the Poisson contribution dominates for a large
survey like the SDSS. For parameters of the full SDSS
survey, roughly 10 million galaxies in each sample and an area of 
10,000 square degrees, the statistical errors are tiny. The errors
are of order 1\% of the signal and have been
multiplied by a factor of 10 in Figure 2 to be visible at all. 
Clearly the signal to noise is
high enough for such a dataset that estimates for sub-samples of galaxies 
by type and varying redshift bin are possible, thus allowing for the
possibility of measuring their halo occupation properties in detail. 
The signal to noise can be scaled to other survey parameters relatively easily
on small scales, where shot noise dominates so that the scaling is
$\omega/\delta\omega \propto \sqrt{n_1 n_2}\sqrt{\Omega}$, where $n_1$
and $n_2$ are the number densities of the two samples, and $\Omega$ is the 
survey area. Figure 3 shows that the errors expected from quasar-galaxy
correlation measurements are significantly higher, due to the smaller
number of quasars. Thus if systematic errors can be controlled, 
magnification bias is more effectly measured using galaxy-galaxy correlations. 

The two major source of systematic errors are photometric
calibration and errors in the photometric redshifts that are not well
characterized or highly correlated with determination of the galaxy type. 
The latter leads to contamination of the samples: e.g. a foreground galaxy may
be assigned to the background sample in the galaxy-galaxy case, or may
in reality be physically associated with the quasar. In either case
the auto-correlation signal, which is much stronger than the lensing
signal, will get mixed in. Moessner \& Jain (1998) estimate the required
accuracy for photometric redshifts for a desired accuracy in the 
cross-correlation.  Given the miniscule statistical errors, characterizing the 
systematic errors accurately will be paramount for any interpretation of 
results from either the CFHT or SDSS surveys.

For the quasar-galaxy $\omega(\theta)$ shown in Figure \ref{fig:qg}
the error bars are larger because of the smaller number of quasars
compared to background galaxies. Even so, they are at the 1\% level
and would allow for a definitive measurement from the SDSS provided
systematic errors are under control. 

Figure \ref{fig:halo} shows the cross-correlation for halos of 
different masses. This is not an observable, but it is helpful to
see the wide variation in the signal as a function of halo mass 
since galaxies are modelled as occupying these halos. Clearly small
changes in the $N_{\rm gal}(m)$ relation will change the expected
signal since it will populate halos of different masses with galaxies
differently. The dot-dashed curve shows the prediction for halos
of mass $10^{13}M_\odot$, which have a very large amplitude at small
scales. This is relevant for the SDSS since it will produce a sample
of Large Red Galaxies, which are expected to occupy the centers of
massive halos. A more quantitative connection is beyond the scope of
this paper. 
 
The redshift distribution of galaxies we have used has not been chosen to 
be consistent with the models for $N_{\rm gal}(m)$. This amounts to 
requiring from survey measurements that some (typically small) fraction 
of galaxies be discarded before estimating the cross-correlation to 
ensure consistency. 

\section{Discussion}

We have used the halo model of clustering to estimate the cross-correlation
induced by magnification bias between samples of galaxies and quasars
at different redshifts. Our focus has been on the effect of the
model for galaxies used on the predicted signal. We used fits to the
GIF simulations for $N_{\rm gal}(m)$, the mean number of galaxies for 
halos of mass $m$, to find the number of galaxies as a function of
redshift, magnitude and galaxy type. We find that the 
predicted cross-correlation is very sensitive to these parameters. 
Its amplitude varies on arcminute scales is a factor of 2 larger or
smaller compared to an unbiased population, 
depending on the redshift range and galaxy type of the foreground
population. The reason for this wide range is that on
small scales the relation of the galaxy distribution
to the mass is complex and cannot be described by a constant bias factor
of order unity. Further, with increasing redshift a
given apparent magnitude limit corresponds to 
brighter absolute magnitudes and therefore 
the selection of galaxies with higher $m/m_*$, 
which tend to be more strongly clustered. Our results on arcminute
scales therefore differ from 
previous studies which used linear bias models. 

The state of measurements of qso-galaxy correlations are summarized
in Ben\'\i tez et al (2001) and 
Guimaraes et al (2001). Clearly some of the measurements are affected
by incomplete sampling and other observational systematics. Others find
excess correlations on large scales, which are still difficult to 
reconcile with models. But most anomalous measurements are on arcminute
scales where we find great sensitivity to the redshift distribution
and galaxy type used for the measurement. Thus variations of 
a factor of 2 to even 10 in extreme cases can be explained. In existing 
data, the galaxy samples used are difficult to characterize 
given the lack of redhift or color information. The data
discussed in Ben\'\i tez \& Martinez-Gonzalez (1995) show that the sample with 
red galaxies
does have higher signal -- this is consistent with our results, but it is 
difficult to be quantitative given the partial information available for the 
data. 

In view of the massive improvement in the available sample expected from the
SDSS and other forthcoming surveys such as the CFHT Legacy survey, 
we have chosen to simply provide model predictions for these surveys
rather than attempt a detailed analysis of past measurements. Our models
can be extended significantly, to include more detailed galaxy types
and galaxy models that are checked for consistency with SDSS measurements
of the galaxy auto-correlations. 

The statistical errors expected from the SDSS survey are very small, 
about 1\% of the signal for the full sample from galaxy-galaxy correlations. 
We have focused on scales
below 20 arcminutes, where the signal to noise is high. On these scales
the data from the SDSS can be split by galaxy type and redshift bins 
to make a detailed study of galaxy clustering in relation to the dark
matter distribution. On large-scales the interpretation is simple, giving
the bias parameter of desired galaxy sub-samples as a function of redshift. 

What can we learn from sub-Mpc scale measurements of the 
galaxy-mass correlation? In the halo model, the 1-halo term is a linear 
measure of (the product of) $N_{\rm gal}(m)$ and other parameters:
the halo mass function, halo mass profile and galaxy profile. 
Scranton (2002) analyzed the accuracy with which
different parameters describing the galaxy distribution can be constrained
using angular correlations from the photometric SDSS survey. Our results
show that valuable additional information can be gained using measurements
of magnification bias. In particular, the lensing induced cross-correlation
is a measure of projected galaxy-mass correlations: it is thus linear in 
the parameters of the galaxy distribution such as the mean halo occupation 
number of galaxies $N_{\rm gal}(m)$. Galaxy clustering measurements probe
the second moment of this distribution, and therefore one is required to 
make an assumption of how the second moment scales with the mean (on large
scales galaxy clustering measurements do constrain the first moment 
through the two-halo term, but only averaged over a broad mass range). 
The magnification bias measurements considered here would constrain 
$N_{\rm gal}(m)$ directly and thus provide the basis for interpreting
the higher order moments from galaxy clustering measurements. Further, 
the lensing measurement would help break the degeneracies between the 
various parameters. 

Scranton (2002) gives the accuracy with which more than 10 parameter 
combinations will be constrained by SDSS measurements of the angular 
clustering of galaxies binned by photometric redshift. For the basic
parameters of $N_{\rm gal}(m)$  in Table 1, 
he typically finds accuracies of well below 1\%. For the lensing induced 
cross-correlation, the signal is significantly smaller and 
for the dominant parameters in $N_{\rm gal}(m)$ such as $\alpha$ 
one can expect better than 10\% level accuracy from the lensing 
measurements alone. 
The main power however will be in combining the lensing and galaxy
clustering measurements. We leave this exercise for future work. 

Our formalism is similar to that of Seljak (2000) 
and Guzik \& Seljak (2001) for galaxy-galaxy lensing. These authors
have used measurements from the SDSS to constrain the mass profiles
of large galaxy halos and the contribution of group halos (Guzik \& 
Seljak 2002; see also McKay et al 2001). 
While in principal magnification bias measures a closely
related cross-correlation (replacing the tangential shear with the
convergence), in practice the measurements have been relevant on different
scales. Galaxy-galaxy lensing has been measured so far on relatively
small scales, the best signal being on $100-200$kpc. It remains to be
seen from the completed SDSS survey and CFHT surveys how the relevant
systematics play out for magnification and shear measures. It will be
of great interest to analyze the measurements jointly to constrain 
galaxy-mass correlations. Similarly the large scale bias parameter 
inferred from shear surveys (Van Waerbeke 1998; Hoekstra et al 2002)
can be directly compared with measurements of magnification bias. 
This will be an extremely useful cross-check since the errors in 
both measurements are  likely to be dominated by systematic errors. 

We have used the weak lensing approximation in this paper, setting
$\mu = 1 + 2 \kappa$. There are corrections to this relation if $\kappa$
or $\gamma$ are of order unity (e.g. M\'enard et al 2002). 
Correlations using the fully nonlinear magnification relation 
$\mu = 1/[(1-\kappa)^2-\gamma^2]$ can be computed using the halo
model approach developed in Takada \& Jain (2003a,b). It leads
to higher amplitudes for magnification bias on subarcminute 
scales, especially for source redshifts $\gsim 1$ 
(Takada \& Hamana 2003); it will therefore be important to incorporate
the nonlinear effects for deep surveys. 
Further, we have treated galaxies as points within halos, rather than
subclumps with mass profiles. Using the formalism of Sheth \& Jain (2002)
we have estimated the effect and found it to be small for the range
of scales studied in this paper. 

\bigskip

We thank N. Ben\'\i tez, A. Connolly, R. Scoccimarro, A. Szalay 
and M. Takada for helpful discussions. We thank the referee, 
Brice M\'enard, whose comments helped improve and correct the paper. 
Some of the work presented here was begun 
during a summer 2002 workshop at the Aspen Center for Physics. 
BJ is supported by NASA grants NAG5-10923, NAG5-10924 and a Keck 
foundation grant.

\onecolumn 

\newpage

\end{document}